\documentclass[12pt,epsfig]{article}
\usepackage{xcolor}
\usepackage{natbib}
\usepackage[top=1.8cm, bottom=1.8cm,left=3.0cm,right=2.5cm]{geometry}

\usepackage{amsmath}

\usepackage{epstopdf}
\usepackage{subfigure}
\usepackage{graphicx}
\usepackage{amsmath}
\numberwithin{figure}{section}
\numberwithin{equation}{section}
\numberwithin{table}{section}

\usepackage{float,latexsym,float,amsmath,amsthm,amssymb,amsfonts,lineno}
\usepackage{tikz}
\usepackage{hyperref}
\usepackage{longtable}
\usepackage{array}

\usepackage{csquotes}
\usetikzlibrary{arrows,automata}
\usepackage{physics}
\usepackage{parskip}
\usepackage{listings}
\usepackage{moreverb}
\usepackage{verbatimbox}
 \usepackage{titlesec}
\usepackage{authblk}
 
\date{}
\title{Mathematical modelling of nutrient-dependent biofilm growth on medical implants}

\author[1,2]{Parna Mandal\thanks{Corresponding author: \href{mailto:parna.pm@outlook.com}{parna.pm@outlook.com}}}
\author[1]{Nigel J. Mottram}
\author[2,3]{Sean McGinty}

\affil[1]{School of Mathematics and Statistics, University of Glasgow, University Place, G12 8QQ, UK}
\affil[2]{Division of Biomedical Engineering, James Watt School of Engineering,
University of Glasgow, University Place, G12 8QQ, Glasgow, UK}
\affil[3]{Glasgow Computational Engineering Centre, James Watt School of Engineering,
University of Glasgow, University Place, G12 8QQ, Glasgow, UK}
\begin{document}
\maketitle

\begin{abstract}
Biofilm infections on medical implants are difficult to eradicate because insufficient nutrient availability promotes antibiotic-tolerant persister cells that survive treatment and reseed growth. Existing mathematical models usually omit nutrient-dependent phenotypic switching between proliferative and persister states. Without this mechanism, models cannot capture how environmental conditions control the balance between active growth and dormancy, which is central to biofilm persistence. We present a continuum model that couples nutrient transport with the dynamics of proliferative bacteria, persisters, dead cells, and extracellular polymeric substances. The switching rates between proliferative and persister phenotypes depend on local nutrient concentration through two thresholds, enabling adaptation across nutrient-poor, intermediate, and nutrient-rich regimes. Simulations show that nutrient limitation produces a high and sustained proportion of persister cells even when biomass is reduced, whereas nutrient-rich conditions support reversion to proliferative growth and lead to greater biomass. The model also predicts that persister populations peak at times that vary with nutrient availability, and these peaks coincide with turning points in biofilm growth, identifying critical intervention windows. By directly linking nutrient availability to phenotypic switching, our model reveals mechanisms of biofilm persistence that earlier models could not capture, and it points toward strategies that target nutrient-driven adaptation as a means to improve the control of implant-associated infections.
\end{abstract}

{\bf Keyword:} Bioflim; nutrient-dependent;  medical implant; persisters; resilience

\section{Introduction}\label{intro}
Biofilms are structured microbial communities that adhere to surfaces and are embedded within a self-produced extracellular matrix. They are a key survival strategy for bacteria, enabling colonisation and persistence in diverse environments, from soil and aquatic systems to industrial equipment, the human body, and medical devices \citep{d2018invasion,acemel2018computer, martin2017new,karunakaran2011biofilmology,wang2010review,costerton1995microbial}. While multi-species biofilms are commonly found in natural habitats, single-species biofilms are more often associated with clinical contexts, particularly on implants,  where they can lead to long-term infections \citep{wallace2016effect}.

The process of biofilm development typically progresses through four stages: initial surface attachment, production of extracellular polymeric substances (EPS), maturation into a structured community, and eventual detachment. The resulting structures confer bacteria with protection against environmental stresses, including antimicrobial agents, making biofilm-associated infections difficult to eradicate. This challenge is particularly pronounced in medical settings, where implant-related infections account for over a quarter of all healthcare-associated infections \citep{sharma2023microbial,arciola2018implant}, posing significant risks such as chronic inflammation, device failure, and the need for surgical revision.

A major reason for the persistence of biofilm infections lies in their ability to withstand antibiotic treatment. This resilience is driven by two distinct mechanisms: resistance and tolerance \citep{davies2003understanding,lewis2001riddle,olsen2015biofilm,uruen2020biofilms}. Resistance involves heritable genetic changes—such as mutations or the acquisition of resistance genes—that enable bacteria to grow despite antibiotic presence. These changes may trigger efflux pump activation, enzyme secretion to inactivate drugs, or other defensive responses. In contrast, tolerance is a transient, non-genetic adaptation. Tolerant cells, particularly persister cells, survive antibiotic exposure without replicating. Once treatment ends, depending on the environmental conditions, they may reactivate and contribute to the growth of the biofilm. Persister cells are typically located in nutrient-deprived regions of the biofilm and are protected by the EPS matrix, which also supports communication through quorum sensing and localised gene expression changes.

Current clinical strategies, including systemic antibiotics, antibiotic-loaded  cements, and surgical debridement, often fail to fully eliminate biofilm infections \citep{knight2019mathematical,dacsbacsi2016mathematical}. In particular, systemic administration often does not achieve adequate antibiotic concentrations at the infection site, especially for bacteria embedded in biofilms or residing intracellularly \citep{arciola2018implant}.  Antibiotic-loaded cements are often similarly ineffective due to the rapid delivery of the antibiotic \citep{arruebo2010drug,zilberman2008antibiotic}. These limitations highlight the need for more effective, targeted treatment approaches.

To support the development of such strategies, mathematical models have become an important tool for studying biofilm behaviour. These models help integrate complex biological and physical processes and enable prediction of treatment outcomes. Several reviews have explored the extensive literature on research into biofilm dynamics, which include many different perspectives -- biological, mechanical, and mathematical \citep{wang2010review,picioreanu2004advances} -- highlighting the complexity of the field and the importance of cross-disciplinary insights. Many existing models have focused on biofilm growth and genetic resistance mechanisms \citep{roberts2019mathematical,dacsbacsi2016mathematical}. However, tolerance driven by phenotypic switching, particularly the reversible transition between proliferative and persister states, has received comparatively little attention.

Some progress has been made in incorporating phenotypic heterogeneity into biofilm models. For instance, \cite{miller2014mathematical} introduced transitions between cell states but did not account for the role of environmental cues, such as nutrient availability, in driving these transitions. Experimental studies have demonstrated that nutrient limitation can trigger the formation of persister cells \citep{shaikh2023effect}, suggesting that models that omit this factor miss a critical aspect of biofilm resilience.

\subsection{Existing mathematical models of biofilm dynamics}\label{lr}
Mathematical models are widely used to study biofilm dynamics and have become essential for investigating growth behaviour. Their importance has been emphasised in numerous reviews that evaluate how such models contribute to understanding biofilm development, structural organisation, and therapeutic response \citep{wang2010review,picioreanu2004advances,davey2000microbial,costerton1995microbial}. A broad spectrum of modelling approaches has been developed, varying in both complexity and scope. These include simplified continuum models in low dimensions \citep{wanner1996mathematical}, spatially oriented frameworks based on diffusion-limited aggregation \citep{fujikawa1994diversity,tolman1989cluster}, models that integrate discrete and continuum scales \citep{ben1994generic}, and more comprehensive systems that couple biofilm growth with fluid dynamics \citep{picioreanu2000effect,dillon1996modeling}.

Early biofilm models focused on nutrient uptake using a one-dimensional spatial representation, where microbial growth was assumed to depend on local nutrient concentration following Monod-type dynamics \citep{atkinson1974overall}. These models helped establish the idea that nutrient diffusion limits bacterial growth in deeper layers of the biofilm, laying a foundation for understanding spatial heterogeneity in biofilms. However, they did not account for structural complexity or differences in bacterial phenotypes. An important extension of this work examined the relationship between nutrient availability and biofilm thickness at steady state \citep{rittmann1980model}. This study showed that there is a minimum bulk nutrient concentration, below which biofilm growth cannot be sustained because the nutrient flux to the biofilm becomes zero. In other words, unless the external nutrient level exceeds this critical threshold, the biofilm will not grow. This result highlighted the essential role of nutrient supply in maintaining biofilm development. Later models introduced a multi-species framework using a continuum approach to simulate microbial dynamics and nutrient transport over time through mass conservation and reaction–diffusion equations \citep{wanner1986multispecies}. These models also incorporated biological processes like biomass degradation and sloughing due to shear stress, offering a more realistic depiction of biofilm growth. However, they often assumed uniform biomass and EPS distribution, which may not reflect the heterogeneity observed at the microscale. Further developments included a more flexible mixed-culture model that allowed the transport of both dissolved and particulate substances within the biofilm matrix \citep{wanner1996mathematical}. This model considered how particle diffusion affects the local liquid content of the biofilm and was validated using experimental data. While it could simulate both short- and long-term growth behaviour, it still relied on a one-dimensional structure and did not fully capture the spatial or phenotypic complexity of real biofilms.

 Cellular Automaton (CA) models and hybrid discrete-continuous approaches have been widely used to simulate biofilm growth, showing strong alignment with experimental observations \citep{pizarro2004two,picioreanu1998new,picioreanu2004advances}. These methods incorporate key biological mechanisms such as bacterial division, intercellular signalling, and nutrient transport. Despite their strengths, a notable drawback of these approaches is the stochastic nature of biomass redistribution, which can introduce algorithmic artefacts and reduce biological interpretability. To address this, deterministic models have been proposed as an alternative, offering improved consistency and reduced computational bias in representing biofilm dynamics \citep{eberl2001new,cogan2004role,alpkvista2007multidimensional}.

Partial differential equation-based continuum models have been extensively used to simulate the spatiotemporal dynamics of biofilms. One such model incorporates density-dependent diffusivity to study how nutrient availability affects biofilm structure. Under nutrient-limited conditions, the model predicts a heterogeneous spatial pattern, whereas nutrient-rich environments produce more compact and interconnected biofilms \citep{eberl2001new}. Another continuum approach models biofilm expansion in a stationary liquid environment by treating the biofilm as a viscous, incompressible fluid, with velocity described using Darcy’s law \citep{klapper2002finger}. A related framework conceptualises the biofilm as a gel made up of water and EPS, where movement arises due to forces on the EPS and the fluid phase, leading to swelling and biomass redistribution \citep{cogan2004role}. Although these models provide insights into structural changes and flow behaviour, they often rely on simplifications such as uniform EPS properties and constant fluid viscosity, which may not reflect biological variability. To address structural heterogeneity more explicitly, a hybrid model was introduced that treats the EPS matrix through a continuum formulation while representing microbial cells as discrete agents \citep{alpkvist2006three}. This multiscale approach enables more accurate analysis of microscale biofilm architecture. Additional developments extended these ideas by formulating systems of equations that couple nutrient transport with biomass dynamics in space and time \citep{alpkvista2007multidimensional}. These models provide a more realistic description of how local nutrient levels influence microbial growth and spatial distribution within the biofilm.  Reaction–diffusion frameworks have also been used to simulate nutrient transport and bacterial consumption, particularly in one-dimensional models that capture nutrient gradients and uptake patterns \citep{beyenal2005modeling}.  Beyond transport and growth, biofilm mechanical properties have been examined through three-dimensional viscoelastic models. These models simulate biofilm detachment and deformation in response to physical forces, providing structural insights that are not captured in simpler representations \citep{towler2007model,bol20093d,bol2013recent}. However, the complexity of such models necessitates detailed parameterisation and empirical validation, which can limit their application across diverse biofilm systems.

The interface between the biofilm and the surrounding fluid is a dynamic boundary governed by attachment and detachment processes. Various modelling approaches have been used to study how these processes affect the structure and thickness of biofilms over time \citep{el2013modeling,d2016qualitative,d2011qualitative}. Detachment is often represented as a function of biofilm thickness or local shear forces, capturing its crucial role in shaping mature biofilm architecture \citep{eberl2006mathematical,wanner1986multispecies,abbas2012longtime}. For instance, elevated shear stress or increased fluid velocity has been shown to impair nutrient transport and promote thinning of the biofilm layer \citep{mavsic2010measuring}. Additionally, the application of antimicrobial agents—particularly at high concentrations—can lead to a reduction or even complete removal of biofilm boundaries, occasionally producing non-unique steady-state outcomes \citep{szomolay2008analysis}. These observations reflect the sensitivity of biofilm development to both physical and chemical external pressures. Nonetheless, much of the modelling literature to date has concentrated on biofilms in later stages of growth, where detachment plays a dominant role in structural evolution \citep{d2018invasion,d2018moving,revilla2016integrated,mavsic2010measuring,szomolay2008analysis,ayati2007multiscale}.

Additional modelling studies have extended existing frameworks or developed new ones to investigate how different environmental or biological factors affect biofilm boundaries \citep{verotta2017mathematical,wallace2016effect,yazdi2014locomotion,ahmed2012analysis,taherzadeh2012mass,mavsic2010measuring,overgaard2009application,szomolay2008analysis}. One such factor is cell death, particularly in the lower layers of the biofilm adjacent to the substrate. Models have shown that both constant and nutrient-dependent rates of cell death can significantly influence the final steady-state thickness of the biofilm \citep{wallace2016effect}. As nutrient availability declines, increased cell death in these basal layers contributes to internal thinning, in addition to external detachment. Notably, as cell death rates rise, a corresponding reduction in biofilm thickness is observed. While valuable, these models primarily address nutrient regulation in the context of cell mortality, rather than broader phenotypic transitions such as the formation of persister cells. As such, they illustrate an important but narrow application of nutrient-dependent modelling. This highlights a broader gap in the literature surrounding nutrient-driven mechanisms of biofilm tolerance, an area that the current study aims to explore in greater depth.

While considerable progress has been made in understanding biofilm behaviour, several limitations continue to impede the development of effective therapies for implant-associated infections. One such limitation is the lack of attention to biofilm growth under non-antibiotic conditions. Most modelling studies have concentrated on biofilms exposed to antimicrobial agents, resulting in a limited understanding of how biofilms naturally establish and evolve in stress-free environments. This gap makes it challenging to differentiate between baseline biofilm characteristics and those altered by antibiotic exposure, thereby hindering accurate evaluation of treatment effects.

A further shortcoming in current biofilm models is the inadequate representation of bacterial phenotypes such as proliferative, persister, and dead cells, especially in relation to how these states shift in response to nutrient availability. These phenotypes are central to biofilm robustness and structure, particularly when subjected to antibiotic stress. Although some models have attempted to capture phenotypic switching dynamics \citep{miller2014mathematical}, they often omit nutrient-dependent regulatory mechanisms that are known to drive these transitions. Nutrient levels are a key modulator of bacterial metabolism and can influence whether cells continue growing, enter dormancy, or adopt survival strategies under stress. Experimental work has demonstrated that nutrient starvation can lead to persister formation \citep{amato2014nutrient,prax2014metabolic}, making it important for models to integrate such nutrient-driven behaviour. Persister cells, often embedded deep within the extracellular matrix, contribute significantly to the resilience of biofilms in implant-related infections. Despite recognition of phenotypic diversity in some models, the absence of nutrient-regulated transitions, particularly in the context of antimicrobial exposure, remains a substantial limitation. Under nutrient-deprived conditions, biofilms may activate tolerance mechanisms such as enhanced EPS production and increased formation of persister cells. The inability to capture these dynamics restricts current modelling frameworks from fully describing biofilm responses.

To address these limitations, we present a new mathematical model that incorporates nutrient-driven phenotypic switching between proliferative and persister cells. By doing so, the model aims to provide a more complete representation of tolerance and improve understanding of the factors that sustain biofilm infections, particularly in early-stage development. This framework extends the biofilm model developed in \citep{miller2014mathematical} by incorporating a key regulatory mechanism: the effect of nutrient availability on transitions between proliferative and persister states. By introducing nutrient-driven phenotypic switching, the model allows for an investigation into how spatial variations in nutrient concentration influence the emergence of persister cells and their potential reversion to active growth, a process that plays a key role in bacterial tolerance and has important implications for understanding the structure and persistence of biofilms, particularly under stress conditions where nutrient gradients are pronounced. The insights gained from such a model have the potential to help inform  treatment strategies, such as the optimal delivery method of antibiotics.

\section{Mathematical modelling}\label{ms}
We develop a mathematical model to investigate biofilm dynamics by explicitly representing spatial and temporal variations in bacterial populations, EPS and nutrient concentration. This modelling framework differentiates between bacterial phenotypes, namely, proliferative, persister, and dead cells -- and includes EPS as a key structural component. The model is designed to capture how local nutrient concentrations influence these subpopulations over time.

\begin{figure}[!t]%
\centering
    \includegraphics[width=0.8\textwidth]{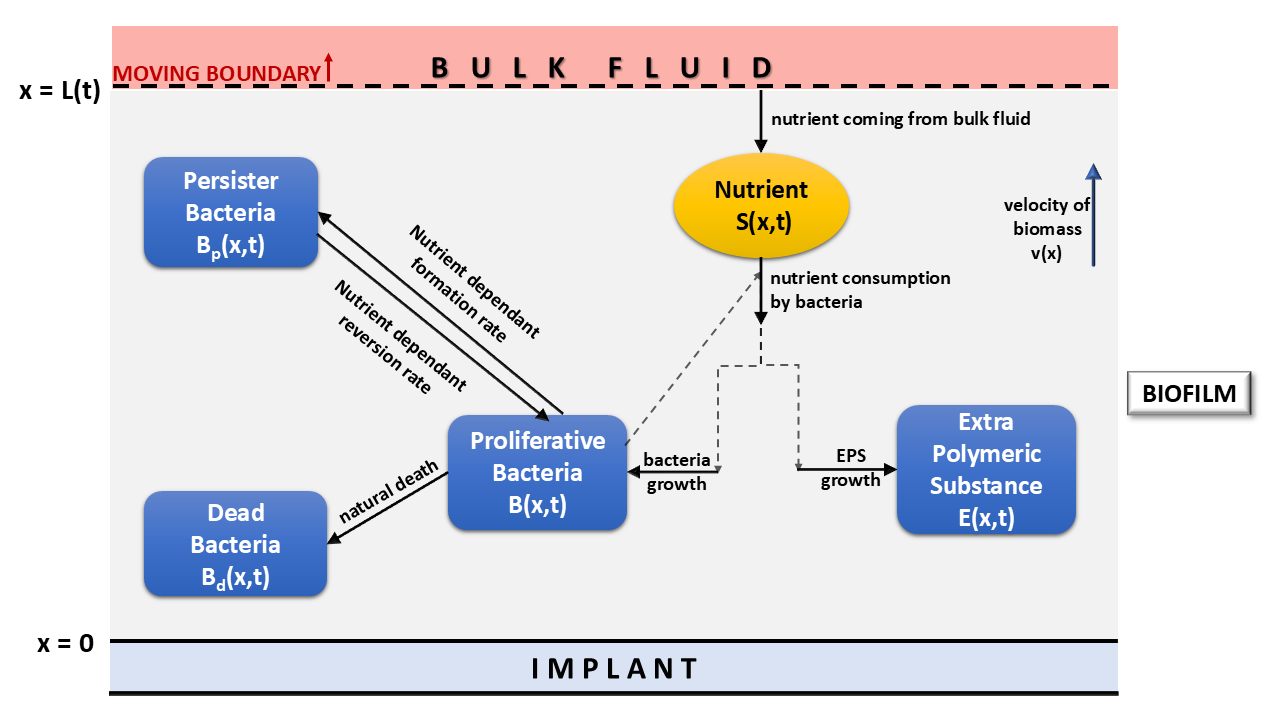}
    \caption{A one-dimensional representation of biofilm development on an implant surface, with spatial position $x$ in the vertical direction. The model domain, $0\le x\le L(t)$, is limited to the biofilm region and does not explicitly include the bulk fluid or the implant medium. Biofilm components are represented by blue boxes and include proliferative bacteria, $B (x,t)$, persister cells,  $B_p (x,t)$, dead cells, $B_d (x,t)$, and EPS, $E (x,t)$. The nutrient concentration, $S (x,t)$, present within the biofilm pores, is indicated by the yellow box. The position $x = L(t)$ denotes the time-dependent moving boundary corresponding to the biofilm thickness and $v(x)$ is the velocity of the biomass.}
    \label{fig:sch}
\end{figure}

As illustrated in Figure~\ref{fig:sch}, the model is constructed within a one-dimensional spatial domain defined along the coordinate \( x \), which extends from \( x = 0 \) to \( x = L(t) \), where \( L(t) \) represents the biofilm thickness at time \( t \). The location \( x = 0 \) corresponds to the interface with the implant, which is treated as a fixed boundary. Beyond the biofilm, for \( x > L(t) \), lies the bulk fluid, which supplies nutrients to the biofilm system. This one-dimensional simplification, reducing the inherently three-dimensional geometry to a single spatial axis, allows the model to preserve key biological dynamics while improving mathematical tractability~\citep{zhao2012theoretical, abbas2012longtime}. At the fixed implant boundary \( x = 0 \), there is no movement of the bacterial phenotypes, EPS, and nutrients into the implant region. Additionally, the advective velocity of the biomass, $v(x)$, is set to zero at this boundary. The upper boundary at \( x = L(t) \), by contrast, is a moving interface that evolves over time in response to biomass growth, nutrient availability, and other dynamic factors. Nutrients are supplied from the bulk fluid at this moving boundary, influencing bacterial activity throughout the biofilm.

Nutrient dynamics within the biofilm are governed by transport from the surrounding bulk fluid and by internal advection-reaction-diffusion processes that evolve over time. Here, the term \textit{nutrient} refers to essential solutes such as oxygen or glucose that are required for microbial growth and biofilm development. This transport occurs within the water-filled volume fraction of the biofilm, denoted by \( \phi_{bio} \), which characterises the porosity of the matrix and thereby modulates both diffusion and advection of soluble species.

Nutrient availability within the biofilm is described by the following equation:
\begin{equation}
   \phi_{bio} S_t + \phi_{bio}(vS)_x = \phi_{bio} D_S S_{xx}  - \frac{\mu \phi_{bio} S}{k_S + \phi_{bio} S} B,
   \label{s}
\end{equation}
where \( S(x,t) \) denotes the nutrient concentration, \( v(x,t) \) is the local advective velocity of the biomass, \( D_S \) is the diffusion coefficient, \( B(x,t) \) is the density of proliferative bacteria, and   $()_t$ and $()_x$ denote differentiation with respect to that variable. The terms on the left-hand side of Equation \eqref{s} represent temporal changes and advective transport of nutrients, while the terms on the right-hand side model diffusion and consumption. Nutrient uptake is assumed to occur solely through proliferative bacteria, governed by Monod kinetics with maximum consumption rate \( \mu \) and half-saturation constant \( k_S \). The model assumes that the biofilm does not penetrate the implant surface; therefore, nutrient diffusion into the implant is also considered zero. As a result, a no-flux boundary condition is applied at the base of the biofilm, \( x = 0 \). At the biofilm--bulk fluid interface, \( x = L(t) \), nutrient supply is sustained by the surrounding medium. To represent this, a Dirichlet boundary condition $S=S_0$ is imposed, with $S_0$ denoting the external nutrient availability, ensuring a continuous influx of nutrients into the biofilm from the external environment.The full mathematical expressions of these boundary conditions are provided in Appendix~\ref{apndxA}.

The bacterial community within the biofilm is divided into three phenotypic classes: proliferative bacteria, $B (x,t)$, persister bacteria,  $B_p (x,t)$ and dead bacteria, $B_d (x,t)$. Biomass is assumed to be capable of movement through both advection and diffusion, although its diffusivity is considerably lower than that of nutrients \citep{sankaran2019single}. Proliferative cells play an active role in consuming nutrients to support biomass growth and in the synthesis of EPS, $E(x,t)$. Under unfavourable environmental conditions, such as nutrient limitation, pH shifts, temperature changes, exposure to antibiotics or immune responses, proliferative cells may transition into a dormant persister state. These persister cells exhibit heightened tolerance to antibiotics, contributing to the resilience of the biofilm. In contrast, dead bacteria no longer participate in growth or metabolic activity but continue to occupy physical space in the biofilm.

The dynamics of the proliferative bacterial population are governed by a combination of diffusion, advection, and reaction processes that evolve over space and time. These mechanisms reflect essential biological behaviors: nutrient-facilitated biomass growth, natural cell death, and nutrient-regulated phenotypic switching to and from the persister state. Since nutrient transport occurs within the water-filled volume fraction of the biofilm, \( \phi_{bio} \), all nutrient-dependent interactions are modelled as functions of this parameter.

The governing equation for proliferative bacteria, denoted by \( B(x,t) \), is
\begin{equation}
\begin{aligned}
B_t + (vB)_x &= D_B B_{xx} +  k_B \frac{\mu \phi_{bio} S}{k_S + \phi_{bio} S} B - bB \\
&\quad - \max\left(k_F\frac{S_2-S}{S_2-S_1},0\right) B + \max\left(k_R\frac{S-S_1}{S_2-S_1},0\right) B_p,
\end{aligned}
\label{b}
\end{equation}
where \( D_B \) is the diffusion coefficient of proliferative bacteria. Proliferative growth is driven by nutrient availability and follows Monod kinetics, represented by the second term on the right-hand side of Equation \eqref{b}, while the third term, \( bB \), denotes natural cell death.

\begin{figure}[!t]
    \centering
    \includegraphics[width=0.8\textwidth]{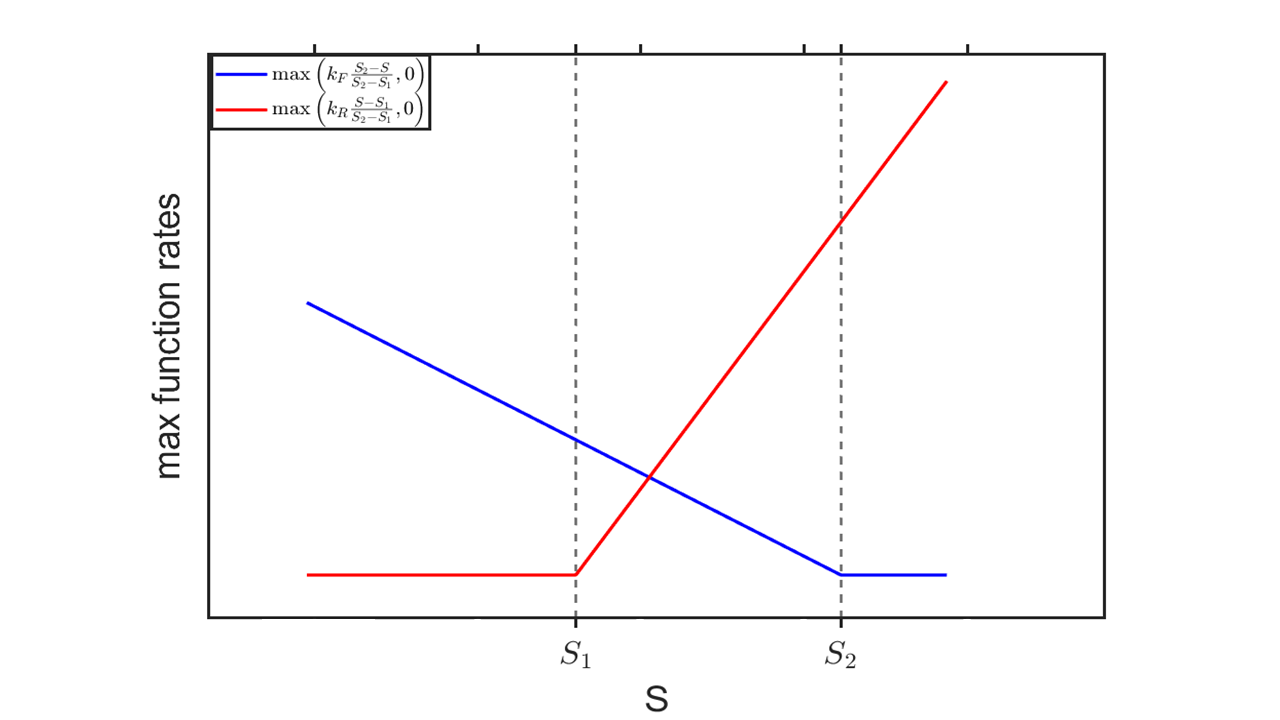}
    \caption{Schematic of the formation and reversion rate of persister bacteria as a function of \( S \).}
    \label{fig:max}
\end{figure}
The final two terms the right-hand side of Equation \eqref{b} model the phenotypic switching between proliferative and persister states and are controlled by two critical nutrient thresholds, \( S_1 \) and \( S_2 \), which define three distinct regimes, as shown in Figure~\ref{fig:max}. These thresholds enable the model to capture how bacterial populations adjust to variations in nutrient availability within the biofilm. When the nutrient concentration drops below \( S_1 \), the local environment is no longer conducive to cell proliferation. Under such nutrient-deprived conditions, the model assumes that proliferative bacteria convert into persisters which is a dormant phenotype with reduced metabolic activity that enhances survival during stress, including nutrient limitation. At this stage, the transition from proliferative to persister is maximised, while reversion is completely suppressed, reflecting an environment unsuitable for active growth. In contrast, when the nutrient concentration exceeds \( S_2 \), conditions become favourable for growth and division. The model therefore assumes that persister cells revert to the proliferative phenotype, with the reversion rate maximised and the formation of new persister cells completely halted. This allows the population to take full advantage of the abundant resources and contribute to biofilm expansion. For nutrient concentrations between \( S_1 \) and \( S_2 \), the model describes a transitional regime. In this intermediate range, both persister formation and reversion occur concurrently, with rates modulated by nutrient availability. Specifically, as nutrient levels increase from \( S_1 \) to \( S_2 \), the formation rate of persisters decreases, while the reversion rate rises.  This formulation allows the model to represent a graded and adaptive bacterial response to spatial and temporal nutrient fluctuations, capturing how local conditions influence both growth and dormancy in biofilm populations.

The model incorporates a spectrum of nutrient concentrations introduced from the bulk fluid, enabling the investigation of how nutrient levels affect biofilm behaviour. While the typical nutrient range is informed by values reported in \citep{miller2014mathematical}, this study extends the analysis by examining three distinct nutrient regimes: nutrient-poor, intermediate, and nutrient-rich. The nutrient-poor case is set below the established average range, the intermediate case lies within that typical range, and the nutrient-rich scenario corresponds to values near the upper bound.

The behaviours of persister cells, dead cells, and EPS are described using the same advection-diffusion-reaction framework applied to proliferative bacteria. Their respective reaction terms are formulated in parallel with those of the proliferative phenotype to maintain consistency across bacterial subpopulations.

The persister population,  \( B_p(x,t) \), is governed by
\begin{equation}
(B_p)_t + (v B_p)_x = D_{B_p} (B_p)_{xx} + \max\left(k_F\frac{S_2 - S}{S_2 - S_1}, 0\right) B - \max\left(k_R\frac{S - S_1}{S_2 - S_1}, 0\right) B_p,
\label{bp}
\end{equation}
where \( D_{B_p} \) is the diffusion coefficient for persister cells. This equation reflects phenotypic switching driven by nutrient concentration: persisters are formed from proliferative bacteria under nutrient-limited conditions (\( S < S_2 \)) and revert back when nutrient availability improves (\( S > S_1 \)). Due to their dormant nature, persisters are not assumed to grow or divide.

The dead bacterial population, \( B_d(x,t) \), evolves according to
\begin{equation}
(B_d)_t + (v B_d)_x = D_{B_d} (B_d)_{xx} + bB,
\label{bd}
\end{equation}
where \( D_{B_d} \) is the diffusion coefficient for dead cells within the biofilm. The source term \( bB \) denotes the natural death of proliferative bacteria. This formulation couples dead cell generation directly to the decline of the proliferative population.

The EPS, \( E(x,t) \), is produced by proliferative bacteria and evolves according to
\begin{equation}
E_t + (vE)_x = D_E E_{xx} + k_E \frac{\mu \phi_{bio} S}{k_S + \phi_{bio} S} B,
\label{e}
\end{equation}
where \( D_E \) is the diffusion coefficient for  within the biofilm. EPS synthesis is modelled using Monod-type kinetics, similar to biomass growth, reflecting its regulation by local nutrient availability. Although EPS primarily functions as a structural matrix with restricted mobility, the diffusion term accounts for gradual redistribution due to internal reorganization or detachment processes within the biofilm \citep{frederick2011mathematical}.

The advective movement of biomass is a critical component of the model, governing how biofilm constituents are redistributed and how the biofilm expands over time. This motion arises from the principle of volume conservation and reflects the balance of internal forces within the system. As biomass accumulates locally, it increases the total volume at that location, which leads to forward displacement of material along the \( x \)-axis and causes the biofilm to grow. This expansion directly links biological growth to physical displacement within the biofilm.
To simplify the system, the model assumes that the biofilm is incompressible. Therefore, the sum of volume fractions of all constituents equals one at every point in space and time:
\begin{equation}
1 = \phi_{bio} + \frac{B + B_p + B_d}{\rho_B} + \frac{E}{\rho_E},
\label{phi}
\end{equation}
where \( \phi_{bio} \) is the water volume fraction and \( \rho_B \) and \( \rho_E \) are the respective biomass and EPS densities. This assumption removes the need to explicitly solve for the pressure field that would otherwise arise in a compressible system, thereby reducing computational complexity. Similar incompressibility assumptions have been widely adopted in continuum biofilm models \citep{miller2014mathematical, wanner1996mathematical}, supporting its use in the present study.

The advective velocity \( v(x,t) \) of the biofilm components governs their transport and is derived from this volume constraint. While a pressure-driven formulation of the velocity field is given by Darcy's law, $v = -\lambda P_x$, this is not needed in the numerical simulations, as the pressure field is not explicitly required. Instead, the model uses a volume-based expression:
\begin{equation}
v_x = \frac{1}{1 - \phi} \left( \frac{k_B}{\rho_B} + \frac{k_E}{\rho_E} \right) \frac{\mu S}{k_S + S} B.
\label{v}
\end{equation}
This conservation equation is derived by summing Equations~\eqref{b}, \eqref{bp}, and \eqref{bd}, each normalised by \(\rho_B\), together with Equation~\eqref{e} scaled by \(\rho_E\). The result is then substituted into Equation~\eqref{phi} to obtain a simplified expression. Additionally, the diffusion terms in Equation~\eqref{v} are neglected on the basis that they are significantly smaller compared to the dominant reaction terms, thus enabling computational simplification without affecting model behaviour.

 The biofilm front, defined by the moving interface \( x = L(t) \), advances at a velocity equal to the biomass flow at this boundary. Thus, biological growth directly drives the physical expansion of the biofilm. As a result, the biofilm thickness \( L(t) \) changes over time due to biomass accumulation at the biofilm–bulk fluid interface, given by:
\begin{equation}
\frac{dL}{dt} = v(L).
\label{L}
\end{equation}

To further simplify the system, the model excludes bacterial attachment and detachment dynamics at the biofilm front, assuming these processes are balanced and do not affect large-scale behaviour. Additionally, effects from shear forces and fluid viscosity are neglected. Because we assume that biofilm does not grow inside the implant, the model applies zero-flux boundary conditions for all bacterial phenotypes and EPS at both the implant-biofilm boundary and the biofilm-fluid interface. In addition, the velocity of biomass is set to zero at \( x = 0 \), representing a fixed boundary at the surface of the implant. 

The model assumes a constant value of \( \phi_{bio} = 0.8 \) for the water volume fraction, a choice that is supported by prior studies \citep{miller2014mathematical,wanner1986multispecies}. Initial values for bacterial phenotypes and EPS are determined using the volume constraint equation, Equation~\eqref{phi}, based on the chosen value of \( \phi_{bio} \). For simplicity, the same initial conditions and diffusion coefficients are applied to proliferative, persister, and dead bacteria, as well as to EPS. This assumption is grounded in biological observations of early biofilm development. During the initial stages, the distribution of cells and matrix material are typically uniform, justifying the use of common diffusion parameters \citep{sutherland2001biofilm}. The EPS matrix, being porous and hydrated, allows similarly sized particles and molecules to diffuse with little variation \citep{sankaran2019single}. Additionally, at the onset of biofilm growth, the concentrations of these components are typically low and homogeneous, and so we therefore assume identical low initial conditions for the components. This assumption of identical initial conditions aligns with  previous models  \citep{crouzet2014exploring,sauer2022biofilm} which indicate that the biofilm does not undergo significant differentiation or structural complexity in the early stages of growth. By adopting this simplification, we ensure numerical stability and computational efficiency without compromising the biological realism of our model. Since we model the biofilm at an early developmental stage, we also assume a very small initial biofilm thickness. Matured biofilms can reach up to 100~ $\mu m$ in thickness, as observed in \citep{scalia2025targeting}, and based on this we estimate the initial biofilm thickness to be 10~ $\mu m.$

The initial and boundary conditions, together with the non-dimensional form of the governing equations, are summarised in Appendix~\ref{apndxA}. In the non-dimensionalisation, biofilm constituents and the nutrient concentration, are scaled by the nutrient half-saturation constant; time is scaled by the death rate of proliferative bacteria; the spatial coordinate by the initial biofilm thickness; and the advective velocity by the product of the death rate and the initial biofilm thickness.

\section{Solution Methodology}\label{sm}
 The non-dimensionalised model is implemented and solved in \textsc{MATLAB}. Starting with the initial values of bacterial phenotypes, EPS, and nutrient concentration, equation~\eqref{v} is first solved to compute the advective velocity \(v\). This velocity is then used in equation~\eqref{L} to determine the biofilm thickness at the next time step. Subsequently, the concentrations of bacteria, EPS, and nutrient are updated accordingly. These updated values replace the previous ones, and the process is iteratively repeated until the solution converges within a specified tolerance.

For the numerical solution, the equations governing bacterial phenotypes, EPS, and biofilm thickness are discretised using the explicit Euler method. The nutrient transport equation, formulated as a boundary value problem, is solved using \texttt{bvp4c}, a built-in \textsc{MATLAB} solver designed for two-point boundary value problems. This solver employs a fourth-order collocation method with adaptive mesh refinement and built-in error control, ensuring accuracy and stability while satisfying boundary conditions at both ends of the domain. We use the default settings of the \texttt{bvp4c} solver, including its default tolerances and mesh parameters. All model parameters used in the numerical implementation are summarised in Table~\ref{tab:parameters}.
\begin{longtable} 
{|m{0.14\textwidth}|m{0.33\textwidth}|m{0.26\textwidth}|m{0.13\textwidth}|}
\hline
\textbf{Parameter} & \textbf{Description} & \textbf{Value} & \textbf{Reference} \\
\hline
\endfirsthead
\hline
\textbf{Parameter} & \textbf{Description} & \textbf{Value} & \textbf{Reference} \\
\hline
\endhead
\small
$D_S$ & Diffusion coefficient of nutrient & $2.97\times 10^{-10}\;\rm m^2/s$ &\citep{miller2014mathematical}\\
\hline
$D_B$ & Diffusion coefficient of proliferative bacteria  & $1.485\times 10^{-12}\;\rm m^2/s$&\citep{sankaran2019single} \\
\hline
$D_{B_p}$ & Diffusion coefficient of persister bacteria  & $1.485\times 10^{-12}\;\rm m^2/s$ & \citep{sankaran2019single}\\
\hline
$D_{B_d}$ & Diffusion coefficient of persister bacteria  & $1.485\times 10^{-12}\;\rm m^2/s$ & \citep{sankaran2019single}\\
\hline
$D_E$ & Diffusion coefficient of EPS & $1.485\times 10^{-12}\;\rm m^2/s$ & \citep{sankaran2019single}\\
\hline
$\rho_B$ & Mass density of bacteria &  $200\;\rm kg/m^3$&\citep{xavier2005biofilm} \\
\hline
$\rho_E$ & Mass density of EPS  & $33\;\rm kg/m^3$&\citep{xavier2005biofilm} \\
\hline
$\mu$ & Maximum nutrient consumption rate  & $1.1111\times10^{-4}\;\rm s^{-1}$&\citep{beyenal2003double} \\
\hline 
$k_S$ & Nutrient consumption half saturation constant & $6.5\times10^{-4}\;\rm kg/m^3$ & \citep{beyenal2003double}\\
\hline
$k_B$ & Metabolic rate to biomass production rate conversion factor & $0.86625$ & \citep{beyenal2003double} \\
\hline
$k_E$ & Metabolic rate to EPS production rate conversion factor & $k_B$&\citep{miller2014mathematical} \\
\hline
$b$ & Proliferative bacteria endogenous death rate  & $1.2031\times10^{-5}\;\rm s^{-1}$&\citep{miller2014mathematical}  \\
\hline
$k_F$ & Rate of persister formation  & $7.2188\times 10^{-7}  \rm s^{-1}$&\citep{miller2014mathematical}  \\ 
\hline
$k_R$ & Rate of persister reversion & $2.4063 \times 10^{-5} \rm s^{-1}$&\citep{miller2014mathematical}  \\ 
\hline
$L_0$ & Initial biofilm thickness  & $10^{-5}\;\rm m$ & Estimated from \citep{scalia2025targeting} \\
\hline
$B_0$ & Initial proliferative bacteria concentration &  $4.4147\;\rm kg/m^3$ & Derived from Equation~\eqref{phi}\\
\hline
$B_{p0}$ & Initial persister bacteria concentration &  $4.4147\;\rm kg/m^3$& Derived from Equation~\eqref{phi} \\
\hline
$B_{d0}$ & Initial dead bacteria concentration &  $4.4147\;\rm kg/m^3$ & Derived from Equation~\eqref{phi}\\
\hline
$E_0$ & Initial EPS concentration  & $4.4147\;\rm kg/m^3$& Derived from Equation~\eqref{phi} \\
\hline
$S_1$ & Nutrient concentration threshold below which all proliferative transforms to persister & $5.2\times 10^{-4}\;\rm kg/m^3$&\citep{miller2014mathematical} \\
\hline
$S_2$ & Nutrient concentration threshold above which all persister transforms to proliferative & $8.45\times 10^{-4}\;\rm kg/m^3$&\citep{miller2014mathematical} \\
\hline
$S_0$ & Nutrient concentration initially and at moving boundary & $3.9-9.75\times 10^{-4}\;\rm kg/m^3$&\citep{miller2014mathematical} \\
\hline
$\phi_{bio}$ & Water volume fraction in the biofilm & $0.8$&\citep{quan2022water} \\
\hline
\caption{Description of model parameters and symbols}
\label{tab:parameters}
\end{longtable} 

\section{Results and discussions}\label{rd}
All results are expressed in non-dimensional terms; however, to maintain clarity in the figures, the overline notation typically used to indicate non-dimensional variables has been omitted. For each species (such as \( B \)), a key metric examined is the total concentration, defined as the integral of its concentration across the one-dimensional spatial domain. For instance, the total amount of proliferative bacteria is given by \( \int_0^{L(t)} B(x,t)\, dx \). This represents an areal concentration, mass per unit area, but for consistency with other system variables, the term `total concentration' is used throughout. The non-dimensional and dimensional time units are very close in value, which makes it straightforward to interpret non-dimensional simulation results in physical time. Unless specified otherwise, all simulations are shown at a final non-dimensional time of approximately 88.36, which corresponds to 85 days in dimensional time. This time point was selected because, while the biofilm had not yet reached steady state, it displayed well-developed features and dynamic behaviour by this time.

\subsection{Effects of varying external nutrient availability}\label{S0}
We will refer to the model described in Section~\ref{ms} as the nutrient-dependent phenotype switch model. In contrast, the nutrient-independent phenotype switch model refers to the model developed in \citep{miller2014mathematical} which assumes constant, i.e.~nutrient-independent, transition rates between proliferative and persister bacterial states. In our analysis, we investigate three scenarios of external nutrient availability, $S_0$: a sufficient nutrient case, where the concentration of nutrients coming from the bulk fluid is higher than both \( S_1 \) and \( S_2 \), represented by solid red lines in subsequent plots; an intermediate nutrient case, where the nutrient concentration lies between \( S_1 \) and \( S_2 \), shown by solid black lines; and a nutrient-poor case, where the concentration is below both thresholds, indicated by solid blue lines. For the nutrient-independent phenotype switch model, these same nutrient scenarios are depicted using dashed lines of the corresponding colours. We primarily focus on the temporal variations of the biofilm components. The corresponding spatial distributions are shown in Fig.~\ref{fig:bbpbdevsxs0scaled} of Appendix~\ref{apndxb}.

\begin{figure}[!t]
  \begin{subfigure}(a)
        \includegraphics[width=0.45\textwidth]{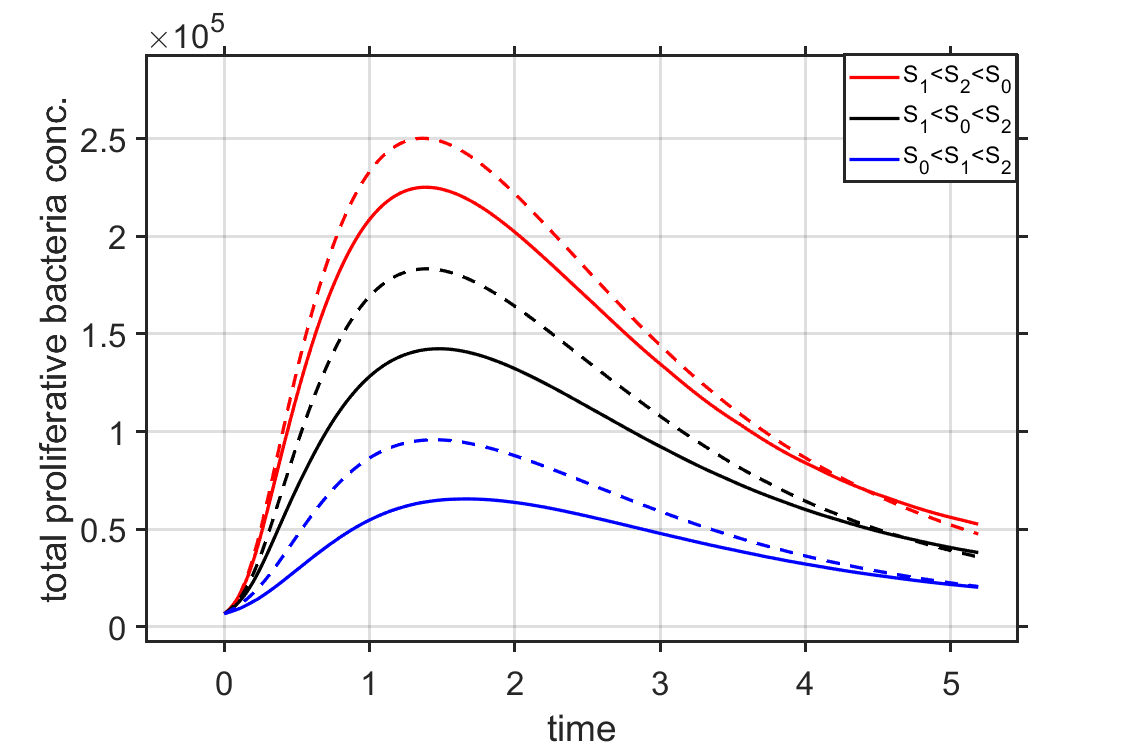}
  \end{subfigure}
   \begin{subfigure}(b)
        \includegraphics[width=0.45\textwidth]{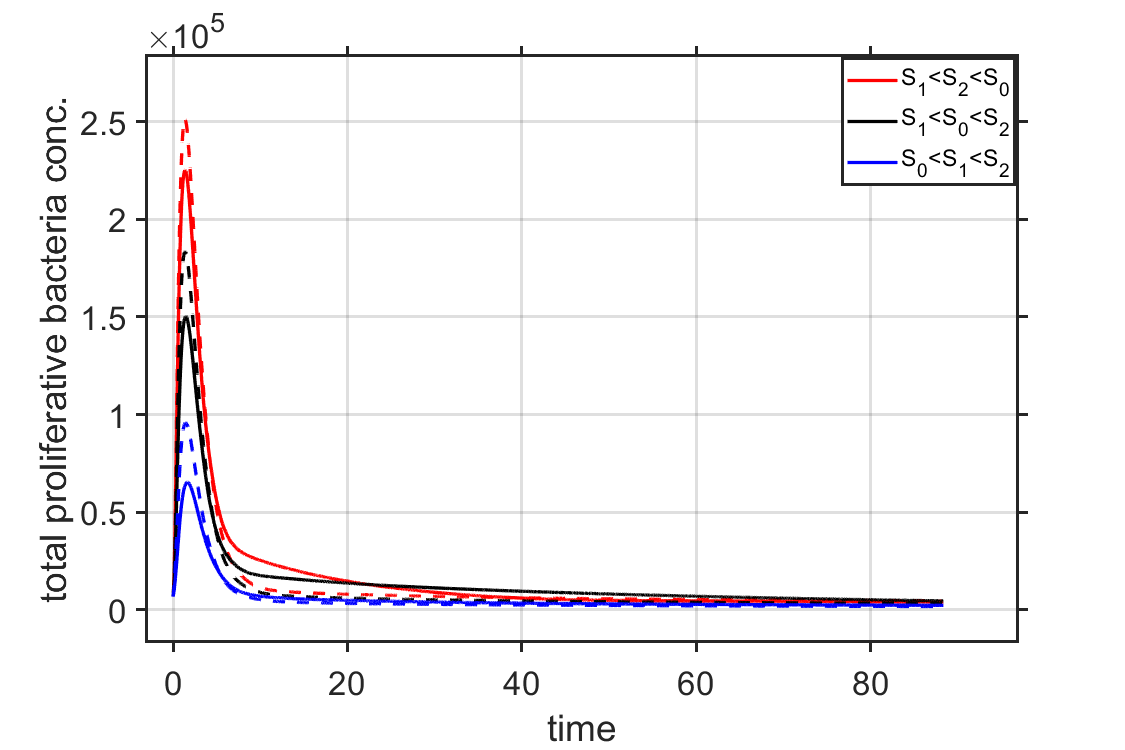}
  \end{subfigure}
   \begin{subfigure}(c)
        \includegraphics[width=0.45\textwidth]{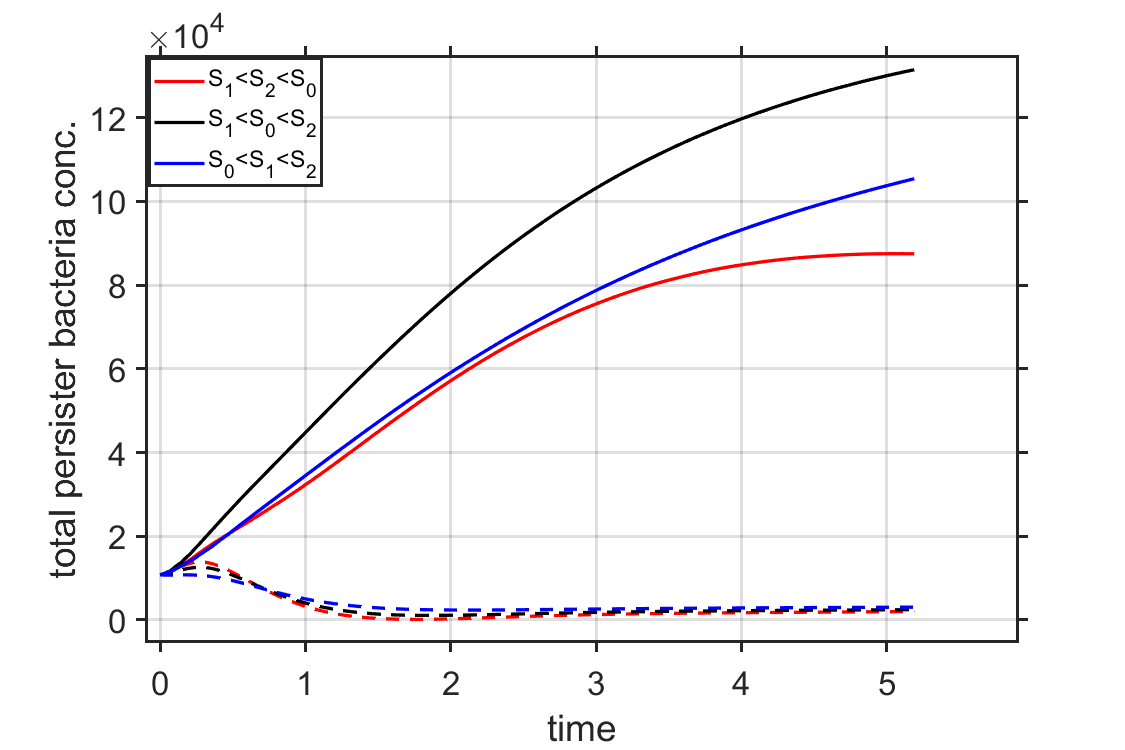}
  \end{subfigure}
   \begin{subfigure}(d)
        \includegraphics[width=0.45\textwidth]{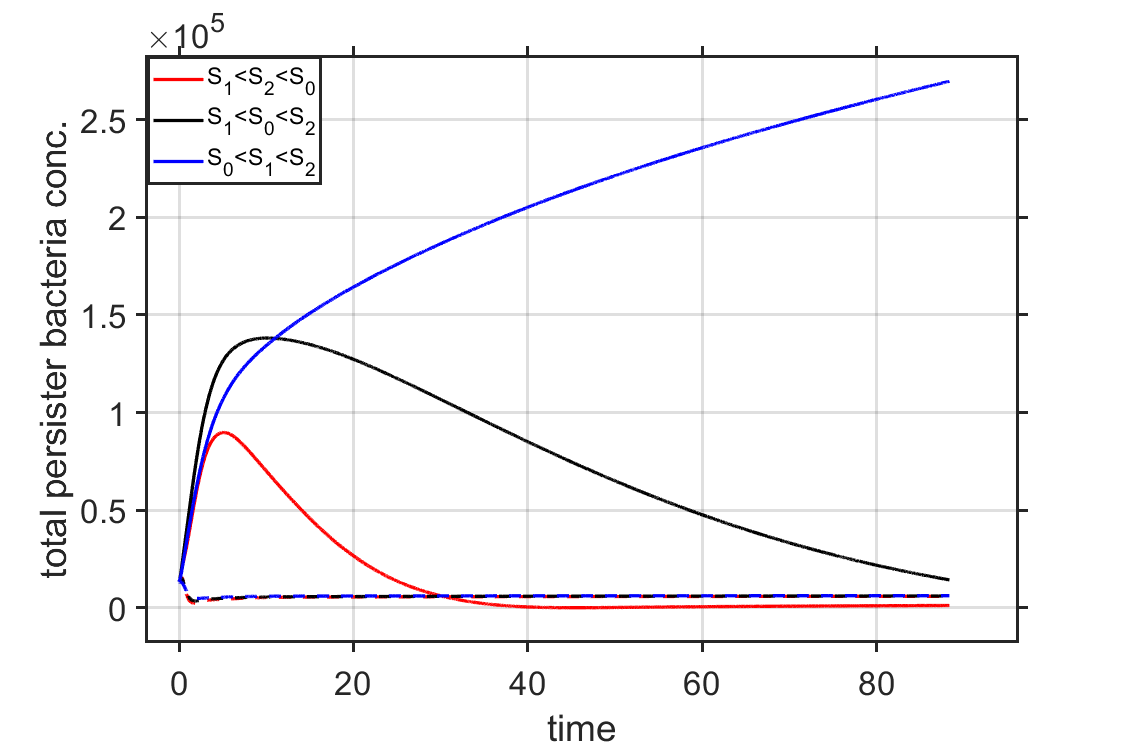}
  \end{subfigure}
   \begin{subfigure}(e)
        \includegraphics[width=0.45\textwidth]{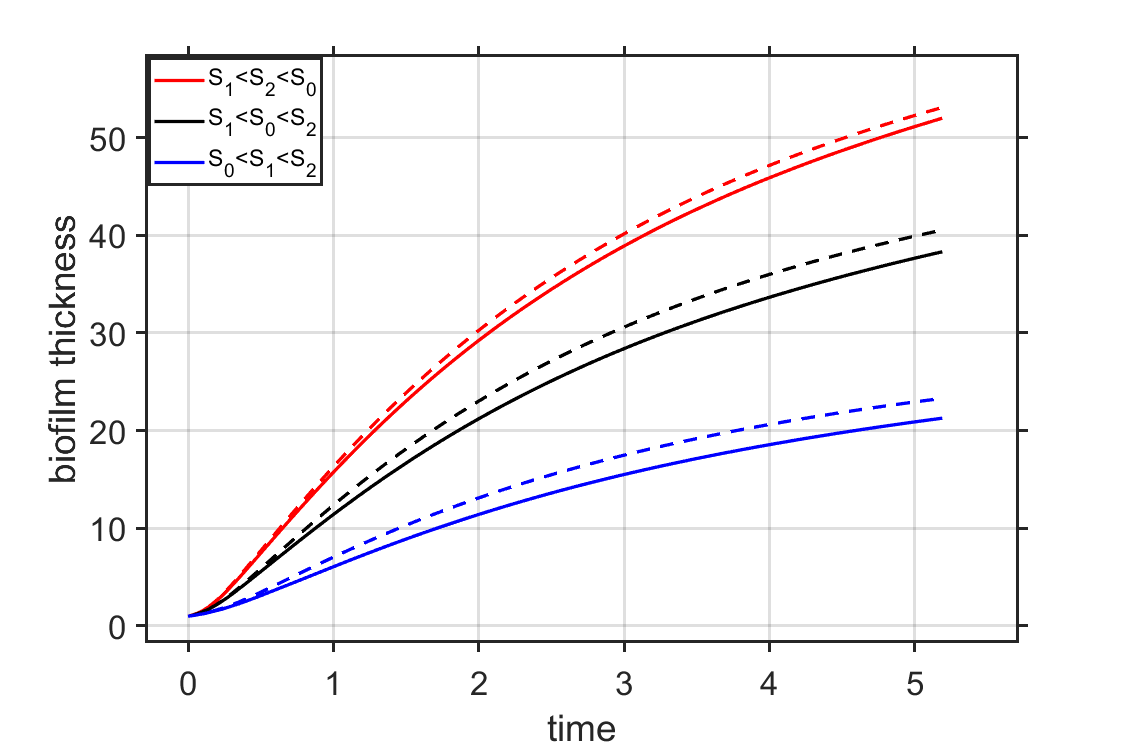}
  \end{subfigure}
   \begin{subfigure}(f)
        \includegraphics[width=0.45\textwidth]{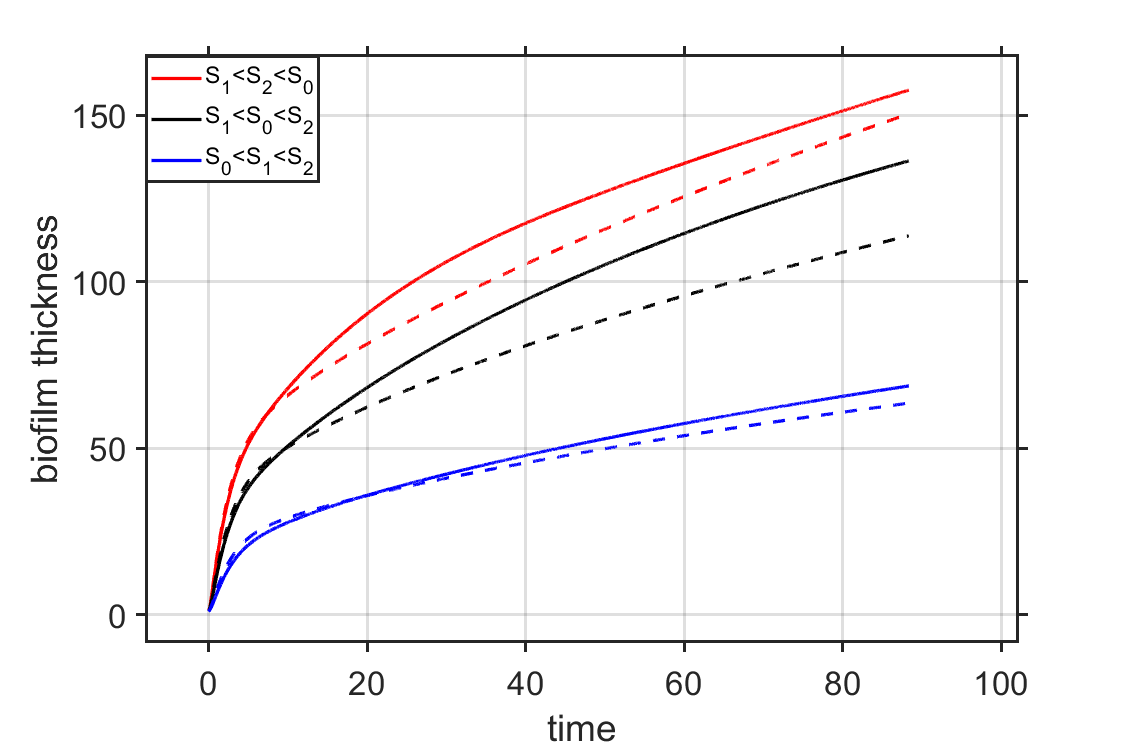}
  \end{subfigure}
 \caption{Effect of varying external nutrient availability, $S_0$, on temporal variation of (a) total proliferative bacteria concentration over 5 days, (b) total proliferative bacteria concentration over 85 days, (c) total persister bacteria concentration over 5 days, (d) total persister bacteria concentration over 85 days, (e) biofilm thickness over 5 days, (f) biofilm thickness over 85 days. The solid lines refer to the nutrient-dependent phenotype switch model, whereas the dashed lines refer to nutrient-independent phenotype switch model. }
     \label{fig:tbbpls0}
\end{figure}

The temporal evolution of biofilm characteristics under varying nutrient concentrations reveals several important trends. As might be expected, biofilm thickness and the total concentration of proliferative bacteria are highest under sufficient nutrient conditions and lowest when nutrients are insufficient. As indicated by dashed lines in Fig.~\ref{fig:tbbpls0}, the nutrient-independent phenotype switch model, shows a higher concentration of proliferative bacteria and greater biofilm thickness up to a non-dimensional time of approximately $5.2$ ($5$ days). However, as time progresses, the biofilm constituents begin to exhibit slower dynamic changes, as seen in Fig.~\ref{fig:tbbpls0}(b) and (f). For proliferative bacteria, the rate of decrease starts to slow down, and for biofilm thickness, the rate of increase becomes more gradual. This behaviour arises as the processes of bacterial growth, death, and phenotypic switching begin to balance each other out, indicating a transition of the biofilm towards a more mature stage. Consequently, the concentration of proliferative bacteria and the resulting biofilm thickness eventually surpass those predicted by the nutrient-independent phenotype switch model.

The behaviour of persister bacteria also varies significantly with nutrient availability. In the early stages for the nutrient-dependent phenotype switch model, persister concentration increases across all nutrient conditions, as shown in Fig.~\ref{fig:tbbpls0}(c). Over time, however, this concentration decreases in both the intermediate and sufficient nutrient scenarios, as illustrated in Fig.~\ref{fig:tbbpls0}(d). The highest persister concentration is observed under nutrient-poor conditions, followed by intermediate and then sufficient nutrient levels. This indicates that although biofilm thickness is reduced in nutrient-deficient environments, the persistence of infection may be more difficult to eliminate due to the elevated presence of persister cells.

A notable distinction between the two models emerges in the behaviour of persister bacteria, seen in Fig.~\ref{fig:tbbpls0}(c) and (d). Under constant transition rates, persister concentration declines because the rate of transition from persister to proliferative exceeds the reverse. In contrast, the nutrient-dependent model shows an initial increase in persister concentration across all nutrient conditions. Even in nutrient-rich environments, not all bacteria immediately revert to the proliferative state, likely due to heterogeneity in bacterial responses or localised nutrient limitations. Over time, the abundance of nutrients supports bacterial growth, resulting in the lowest peak persister concentration among the three conditions. In the intermediate nutrient scenario, the persister concentration rises more substantially than in the nutrient-rich case, reflecting a balance between growth and stress-induced dormancy. Eventually, as nutrient levels become adequate, some persister bacteria revert to the proliferative state, leading to a decline in their numbers.

In Fig.~\ref{fig:tbbpls0}(d) we see that, initially the nutrient-poor condition exhibits the highest persister concentration, followed by intermediate and sufficient nutrient levels. As time advances, the persister concentration in the nutrient-poor environment overtakes that of the intermediate condition. This shift is attributed to prolonged nutrient deprivation, which forces a larger fraction of bacteria to remain in the persister state, ultimately resulting in the highest persister concentration under nutrient-poor conditions. The temporal variation of the dead bacteria and EPS are shown in Fig.~\ref{fig:tbdes0apndx} of Appendix~\ref{apndxb}, as they do not offer additional insight relevant to the primary objectives of this study and are therefore omitted from the main text.

\begin{figure}[!t]
    \centering
    \includegraphics[width=0.7\linewidth]{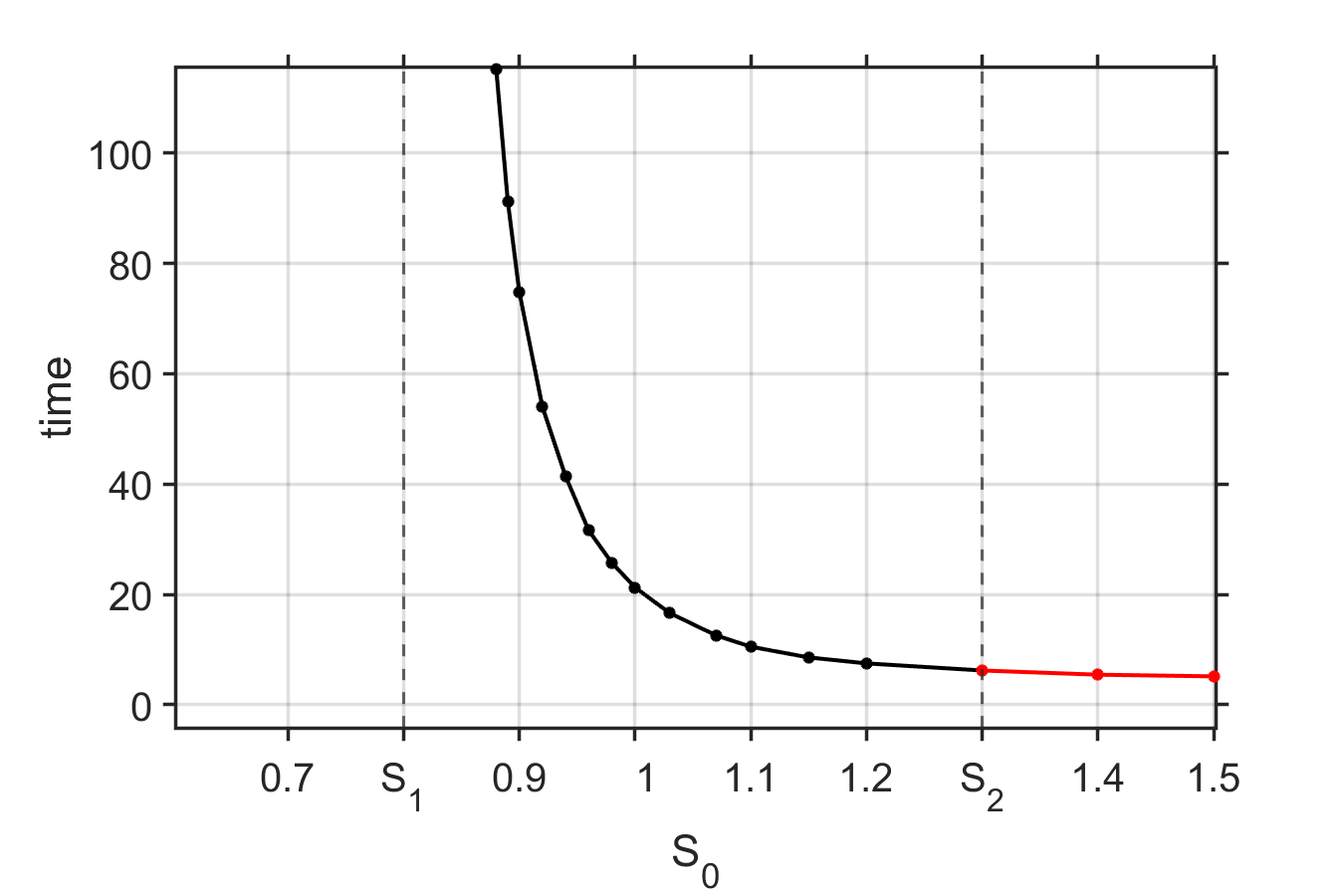}
    \caption{Influence of varying $S_0$ on the times where we can see a peak in total persister bacteria concentration.}
    \label{fig:peaktimes}
\end{figure}

As seen in Fig.~\ref{fig:tbbpls0}(d), the timing of the peak in persister bacteria concentration varies with nutrient availability, with the peak occuring earlier as the nutrient concentration increases. The time at which the peak persister concentration occurs, as a function of nutirient level, is plotted in Fig.~\ref{fig:peaktimes}. For low nutrient levels especially when $S<0.88$, we see the maximum persister bacteria concentration at the final simulation time which in this case is $120$ days, indicating that a distinct peak was not observed within the simulated duration. This suggests that under severely nutrient-limited conditions, persister concentration continues to rise gradually throughout the current simulation window, although it is possible that it may eventually decline at later times, as observed under higher nutrient availability scenarios. The environment in this regime remains consistently unfavourable for proliferative growth, preventing a reversal of the phenotypic state. As a result, most bacteria transition into the persister state to survive, leading to a sustained accumulation without a subsequent decline in persister concentration. A peak in this regime may still emerge, but only over longer time horizons beyond the current simulation window.

Beyond this threshold, a transition in behaviour is observed, the peak time decreases sharply as $S_0$ increases. This trend reflects a more rapid onset of nutrient consumption and subsequent nutrient depletion in environments with higher initial nutrient supply. Greater availability of nutrients initially supports faster proliferation, leading to quicker nutrient exhaustion. This, in turn, induces an earlier and a faster accumulation of persister cells, followed by their eventual decline as nutrient levels recover or stabilise. 

Importantly, the timing of these peaks in total persister bacteria concentration also coincides with a distinct inflection in biofilm growth dynamics. As shown in Fig.~\ref{fig:tbbpls0}(f), biofilm thickness increases rapidly during the initial phase for all nutrient conditions. However, after a certain time, the rate of growth slows down. This transition time aligns closely with the peak times of persister concentration observed in Fig.~\ref{fig:tbbpls0}(d) and detailed in Fig.~\ref{fig:peaktimes}.

This behaviour emerges from the nutrient-dependent switching dynamics encoded in the model. Initially, high nutrient availability supports a predominantly proliferative population, driving rapid biomass accumulation as seen in Fig.~\ref{fig:tbbpls0}(b). As nutrient levels fall below $S_1$ in some parts of the biofilm, a progressively larger fraction of these cells in those area transitions into the persister state, leading to a decline in active growth. Spatial limitations on nutrient diffusion amplify this effect in the parts nearer the biofilm-implant boundary, further shifting the population balance toward dormancy. As a result, while the biofilm continues to grow, its expansion rate diminishes. The structure becomes increasingly composed of metabolically inactive cells that no longer contribute to growth. This internal reorganisation suggests that the biofilm undergoes a qualitative shift in its growth dynamics, potentially reflecting two distinct regimes driven by nutrient availability and phenotypic adaptation.

The temporal window around the peak times of total persister bacteria concentration represents a critical juncture where interventions may be most effective in curbing biofilm growth. Targeting this time, either by modulating nutrient conditions or interfering with phenotypic switching, could enhance the efficacy of treatment strategies aimed at controlling persistent infections. Moreover, these findings underscore the importance of incorporating environment responsive switching mechanisms in mathematical models to capture the full complexity of biofilm behaviour under variable external conditions.

\section{Conclusions and future directions} 
This work presents a novel mathematical model of biofilm growth that, for the first time, explicitly incorporates nutrient-dependent phenotypic switching between proliferative and persister bacterial states. While previous models have considered phenotypic heterogeneity using fixed transition rates, this model captures how local nutrient availability directly governs the dynamic balance between active growth and dormancy within the biofilm.

The model shows that low nutrient environments lead to elevated and sustained persister concentrations, even when overall biofilm mass is reduced—highlighting a key mechanism behind treatment resilience in nutrient-deprived regions. Furthermore, the model identifies distinct peak times in persister concentration that shift depending on external nutrient supply. These peaks correspond with inflection points in biofilm growth and represent potential intervention windows before the biofilm stabilises into a more tolerant structural configuration.

Compared to the nutrient-independent phenotype switch model \citep{miller2014mathematical}, the nutrient-dependent phenotype switch model predicts delayed but ultimately greater biomass accumulation, especially under moderate nutrient conditions, due to adaptive transitions back to the proliferative state as nutrients become available. These findings illustrate how nutrient dynamics fundamentally shape both the spatial structure and temporal evolution of the biofilm, and suggest that therapies targeting nutrient-driven tolerance mechanisms may be more effective than those focused solely on genetic resistance.

Future work could extend the current framework by incorporating antibiotic dynamics and implant-specific mechanisms, enabling the investigation of coupled effects between nutrient-driven phenotypic adaptation and antibiotic response. Additionally, a more detailed study may yield deeper insight into biofilm control strategies and time-specific treatment approaches. Future extensions could also include coupling to bulk fluid dynamics, modelling shear stress and detachment, and exploring multi-dimensional systems to better capture the full physiological complexity of biofilm development and dispersal.



\section*{Funding}
This research was generously supported by the Engineering and Physical Sciences Research Council (EPSRC) under grant number EP/V519984/1.

\bibliographystyle{plain}


\appendix
\section{Detailed equations and calculations}\label{apndxA}

\subsection{Boundary and initial conditions}
The initial conditions, as described in Section \ref{ms}, are given by
\begin{eqnarray}
    S(x,0)&=&S_0, \label{sic}\\
     B(x,0)&=&B_0 e^{\frac{v(x,0)}{D_B}x}, \label{bic}\\
      B_p(x,0)&=&B_{p0} e^{\frac{v(x,0)}{D_{B_p}}x}, \label{bpic}\\
       B_d(x,0)&=&B_{d0} e^{\frac{v(x,0)}{D_{B_d}}x}, \label{bdic}\\
        E(x,0)&=&E_0 e^{\frac{v(x,0)}{D_E}x}, \label{eic}\\
         L(0)&=&L_0, \label{lic}
\end{eqnarray}
where \( v(x,0) \) denotes the velocity of the biomass at \( t = 0 \) and position \( x \), and the initial condition of bacterial phenotypes and EPS is spatially dependent to ensure consistency with the boundary conditions and to represent a non-uniform distribution across the spatial domain.

At the implant-biofilm interface (\( x = 0 \)), the advective velocity is set to zero,
\begin{eqnarray}
   v &=& 0, \label{vbc}
\end{eqnarray}
and zero-flux boundary conditions are imposed on the concentrations of nutrients and biomass components:
\begin{eqnarray}
   D_S S_x &=& 0,\\
   D_B B_x &=& 0,\\
   D_{B_p} ({B_p})_x &=& 0, \\
   D_{B_d} ({B_d})_x &=& 0, \\
   D_E E_x &=& 0. \label{bc1}
\end{eqnarray}

At the moving biofilm–bulk fluid interface, \( x = L(t) \), nutrient concentration follows a Dirichlet boundary condition:
\begin{eqnarray}
   S &=& S_0. \label{sbc2}
\end{eqnarray}

For biomass components, a no-flux condition is applied, accounting for advection at the moving boundary \( x = L(t) \):
\begin{eqnarray}
   D_B B_x - vB &=&0,\\
   D_{B_p} ({B_p})_x - vB_p &=&0,\\
   D_{B_d} ({B_d})_x - vB_d&=&0,\\
   D_E E_x - vE &=& 0. \label{bbpbdebc2}
\end{eqnarray}
A complete list of model parameters is provided in Table \ref{tab:parameters}.

\subsection{Non-dimensionalisation}\label{nd}
To reduce model complexity and minimise the number of parameters, the system is reformulated using a set of non-dimensional variables. This non-dimensionalisation approach, while not identical to those found in previous studies, follows established practices that facilitate the identification of key dimensionless parameters governing the system dynamics. Similar methods have been adopted in earlier works for analysing biofilm behaviour, providing justification for its use here \citep{ward2005modelling,tam2019mathematical}. 

Bioflim constituents, including the nutrient concentration, are non-dimensionalised by the nutrient half saturation constant. Time is non-dimensionalised by the death rate of proliferative bacteria, the spatial dimension by the initial biofilm thickness, and the advective velocity by the product of the death rate and initial biofilm thickness as 
\[
\overline{B} = \frac{B}{k_S}, \quad \overline{B}_p = \frac{B_p}{k_S}, \quad \overline{B}_d = \frac{B_d}{k_S}, \quad \overline{E} = \frac{E}{k_S}, \quad \overline{S} = \frac{S}{k_S}, \quad \overline{v} = \frac{v}{bL_0},
\]
\[
\overline{t} = bt,  \quad \overline{x} = \frac{x}{L_0}.
\]
Using these non-dimensional variables, the governing equations (\ref{b})–(\ref{L}) are transformed into
\begin{eqnarray}
    \overline{B}_{\overline{t}} + (\overline{v} \overline{B})_{\overline{x}} &=& \overline{D}_B \overline{B}_{\overline{x}\overline{x}} + G_1\frac{\phi_{bio} \overline{S}}{1 + \phi_{bio} \overline{S}} \overline{B} - \overline{B} \nonumber \\
    &&- \max\left(\beta_2\frac{\overline{S}_2-\overline{S}}{\overline{S}_2-\overline{S}_1},0\right) \overline{B} + \max\left(\beta_3\frac{\overline{S}-\overline{S}_1}{\overline{S}_2-\overline{S}_1},0\right) \overline{B}_p, \label{bnd} \\
   {(\overline{B}_p})_{\overline{t}} + (\overline{v} \overline{B}_p)_{\overline{x}} &=& \overline{D}_{B_p} ({\overline{B}_p})_{\overline{x}\overline{x}} + \max\left(\beta_2\frac{\overline{S}_2-\overline{S}}{\overline{S}_2-\overline{S}_1},0\right) \overline{B}  -\max\left(\beta_3\frac{\overline{S}-\overline{S}_1}{\overline{S}_2-\overline{S}_1},0\right) \overline{B}_p,\label{bpnd}\\ 
 ({\overline{B}_d})_{\overline{t}} + (\overline{v} \overline{B}_d)_{\overline{x}} &=& \overline{D}_{B_d} ({\overline{B}_d})_{\overline{x}\overline{x}}  + \overline{B},\label{bdnd} \\
   \overline{E}_{\overline{t}} + (\overline{v} \overline{E})_{\overline{x}} &=& \overline{D}_E \overline{E}_{\overline{x}\overline{x}} + G_2 \frac{\phi_{bio} \overline{S}}{1 + \phi_{bio} \overline{S}} \overline{B}, \label{m3end} \\
   \phi_{bio} \overline{S}_{\overline{t}} + \phi_{bio} (\overline{v} \overline{S})_{\overline{x}} &=& \phi_{bio} \overline{D}_S \overline{S}_{\overline{x}\overline{x}} - G_3 \frac{\phi_{bio} \overline{S}}{1 + \phi_{bio} \overline{S}} \overline{B}, \label{snd} \\
  1 &=& \phi_{bio} + \frac{\overline{B} +\overline{B}_p+\overline{B}_d}{\overline{\rho}_B} + \frac{\overline{E}}{\overline{\rho}_E}, \label{phind} \\
   \overline{v}_{\overline{x}} &=& \frac{1}{1 - \phi_{bio}} \left( \frac{k_B}{\overline{\rho}_B} + \frac{k_E}{\overline{\rho}_E} \right) G_3 \frac{\phi_{bio} \overline{S}}{1 + \phi_{bio} \overline{S}} \overline{B}, \label{vnd} \\
   {\overline{L}}_{\overline{t}} &=& \overline{v}(\overline{L}). \label{Lnd}
\end{eqnarray}
The non-dimensional initial and boundary conditions are identical to the dimensional conditions but with overbars applied to every variable and parameter, and the dimensionless parameters for the model are then given as
\[
\overline{D}_B = \frac{D_B}{b L_0^2}, \quad \overline{D}_{B_p} = \frac{D_{B_p}}{b L_0^2}, \quad \overline{D}_{B_d} = \frac{D_{B_d}}{b L_0^2}, \quad \overline{D}_E = \frac{D_E}{b L_0^2}, \quad \overline{D}_S = \frac{D_S}{b L_0^2},
\]
\[
\quad G_1 = \frac{k_B \mu}{b}, \quad G_2 = \frac{k_E \mu}{b}, \quad G_3 = \frac{\mu}{b},
\]
\[
   \quad \beta_2 = \frac{k_F}{b}, \quad \beta_3 = \frac{k_R}{b},\quad \overline{\rho}_B = \frac{\rho_B}{k_S}, \quad \overline{\rho}_E = \frac{\rho_E}{k_S},\]
\[ \quad \overline{B}_0 = \frac{B_0}{k_S}, \quad \overline{B}_{p0} = \frac{B_{p0}}{k_S},\quad \overline{B}_{d0} = \frac{B_{d0}}{k_S},
\overline{E}_0 = \frac{E_0}{k_S}, \quad \overline{S}_0 = \frac{S_0}{k_S},\quad  \overline{L}_0=\frac{L_0}{L_0}=1.
\]
\section{Results and discussions} \label{apndxb}
\subsection{Temporal plots}
    \begin{figure}[!t]
    \centering
  \begin{subfigure}(a)
        \includegraphics[width=0.45\textwidth]{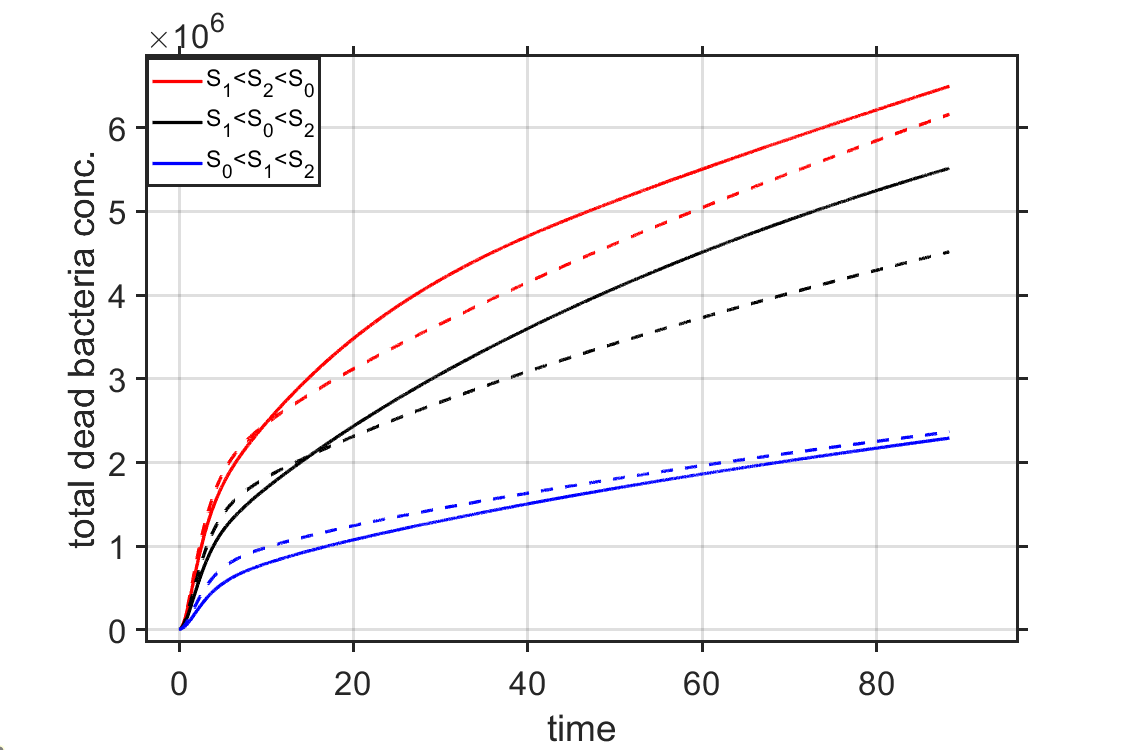}
  \end{subfigure}
   \begin{subfigure}(b)
        \includegraphics[width=0.45\textwidth]{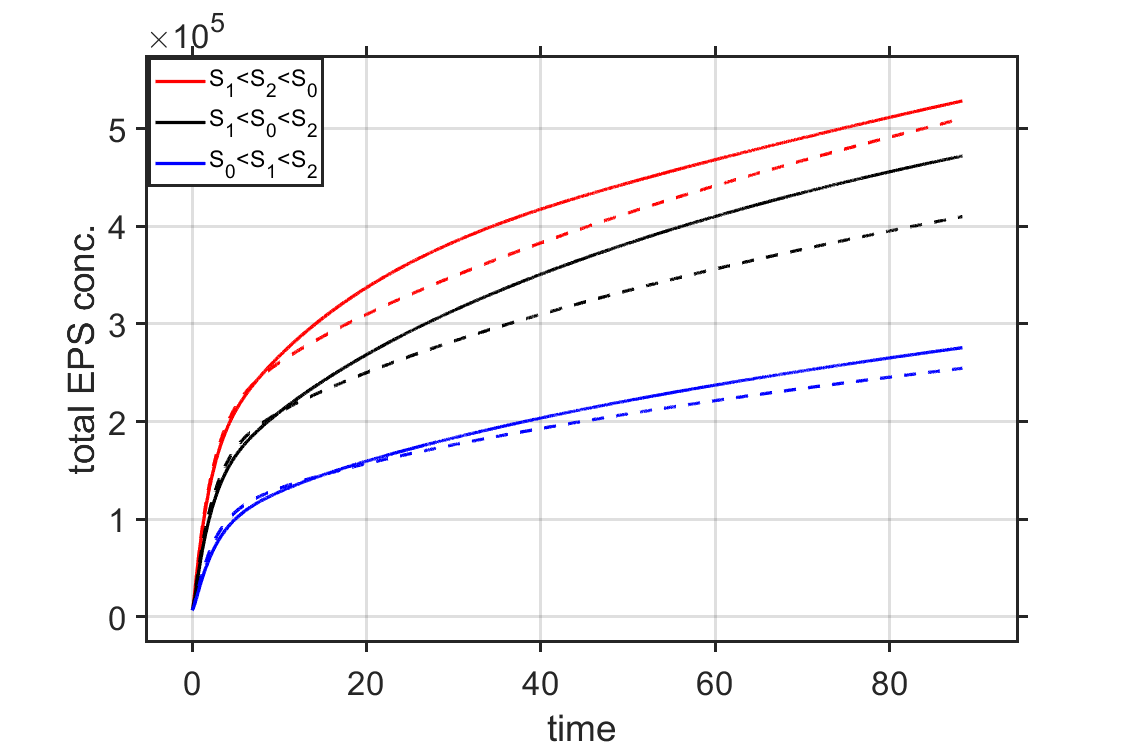}
  \end{subfigure}
 \caption{Effect of varying external nutrient availability ($S_0$) on temporal variation of (a) total dead bacteria concentration over 85 days, (b) total EPS concentration over 85 days. The solid lines refer to the nutrient-dependent phenotype switch model, whereas the dashed lines refer to nutrient-independent phenotype switch model. }
     \label{fig:tbdes0apndx}
\end{figure}

In the nutrient-dependent phenotype switch model, we observe a lower total concentration of dead bacteria and EPS up to a non-dimensional time of approximately $5.2$ ($5$ days) (Fig.\ref{fig:tbdes0apndx}(a) and Fig.\ref{fig:tbdes0apndx}(b), respectively) compared to the nutrient-independent phenotype switch model. This initial reduction is attributed to a lower concentration of proliferative bacteria during this period (Fig.~\ref{fig:tbbpls0}(a)), which leads to reduced cell death and EPS production. As time progresses, the transition dynamics stabilise, enabling sustained bacterial proliferation. Consequently, the total bacterial concentration in the nutrient-dependent phenotype switch model eventually exceeds that of the nutrient-independent phenotype switch model.

The model demonstrates that the highest concentrations of dead bacteria and EPS occur under conditions of sufficient nutrient availability, while the lowest concentrations are observed under nutrient-poor conditions. This outcome is closely linked to the behavior of proliferative bacteria, which respond directly to the surrounding nutrient levels (Fig.~\ref{fig:tbbpls0}(b)). The nutrient-dependent dynamics thus play a critical role in shaping both bacterial mortality and EPS accumulation.
\subsection{Scaled spatial plots}
\begin{figure}[!t]
\centering
  \begin{subfigure}(a)
        \includegraphics[width=0.45\textwidth]{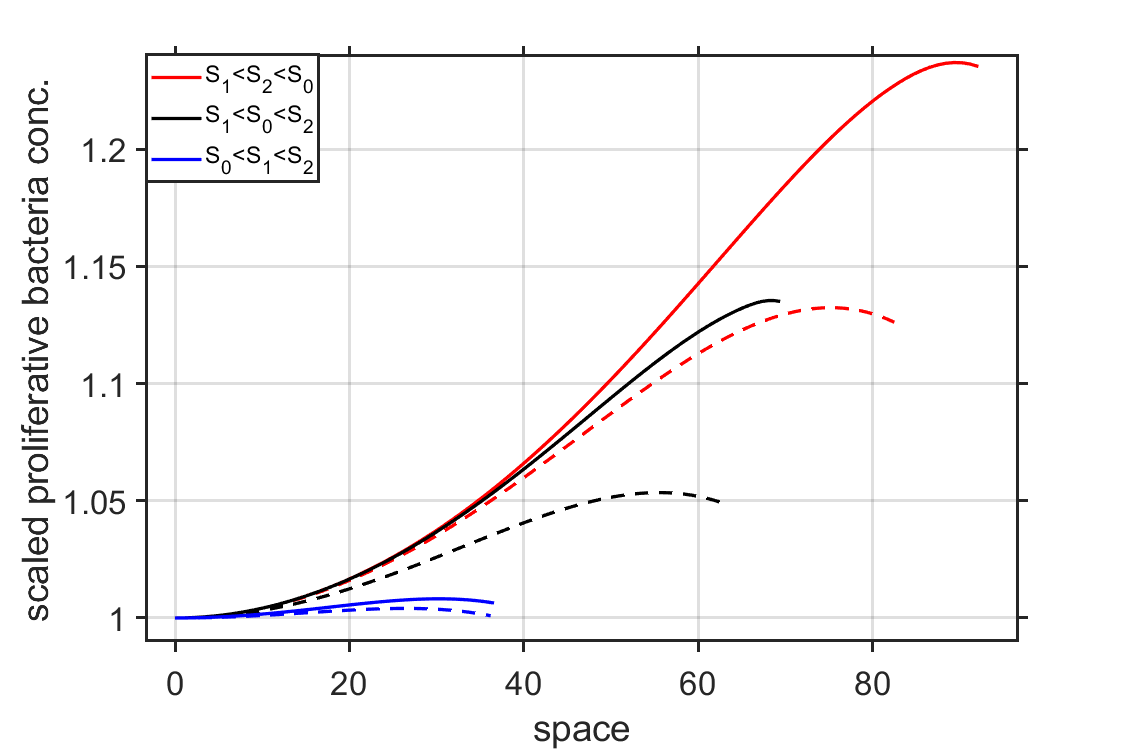}
  \end{subfigure}
   \begin{subfigure}(b)
        \includegraphics[width=0.45\textwidth]{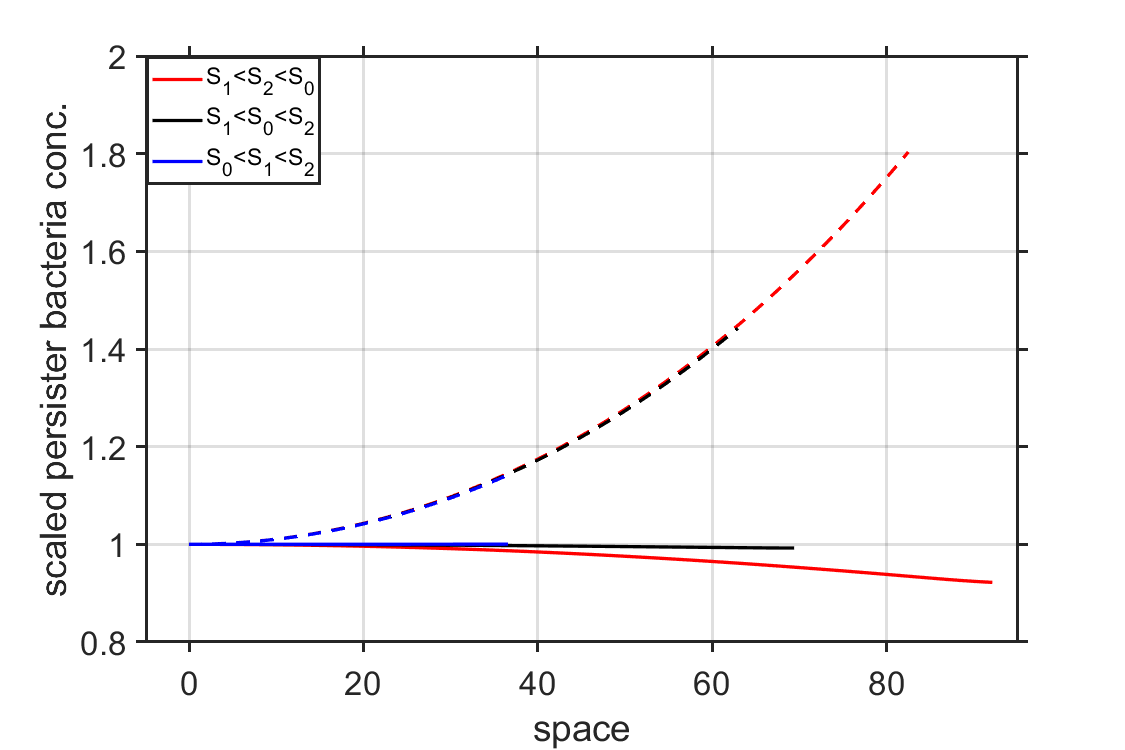}
  \end{subfigure}
  \begin{subfigure}(c)
        \includegraphics[width=0.45\textwidth]{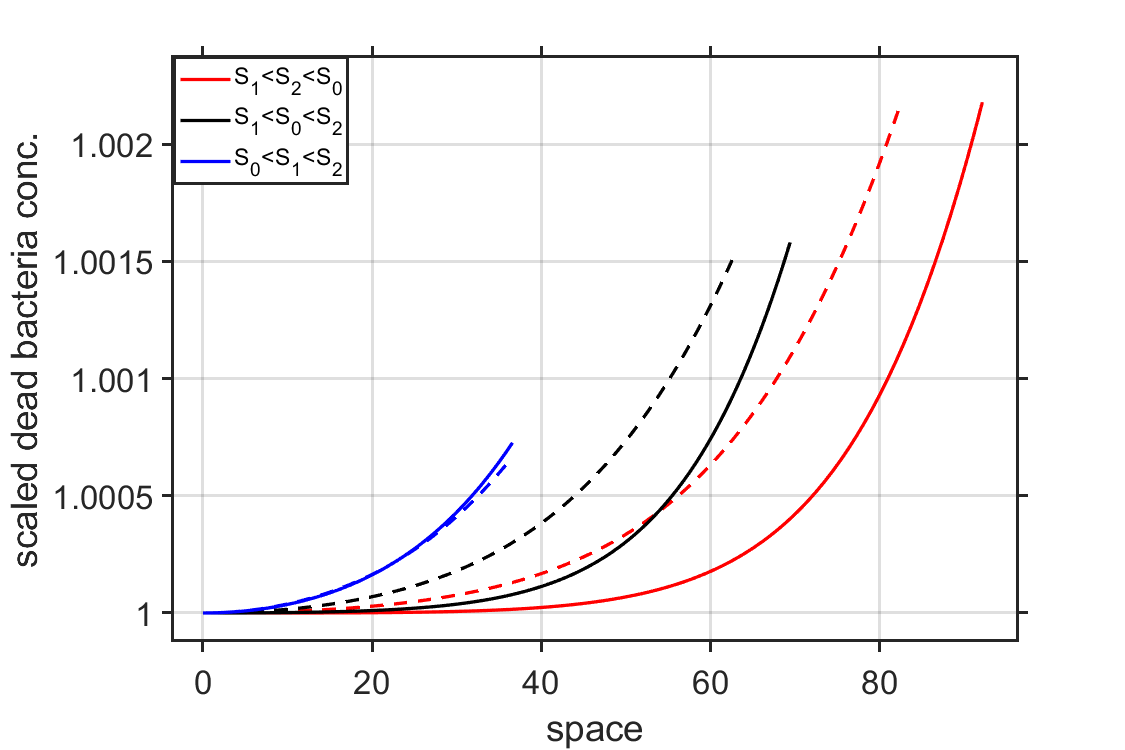}
  \end{subfigure}
  \begin{subfigure}(d)
        \includegraphics[width=0.45\textwidth]{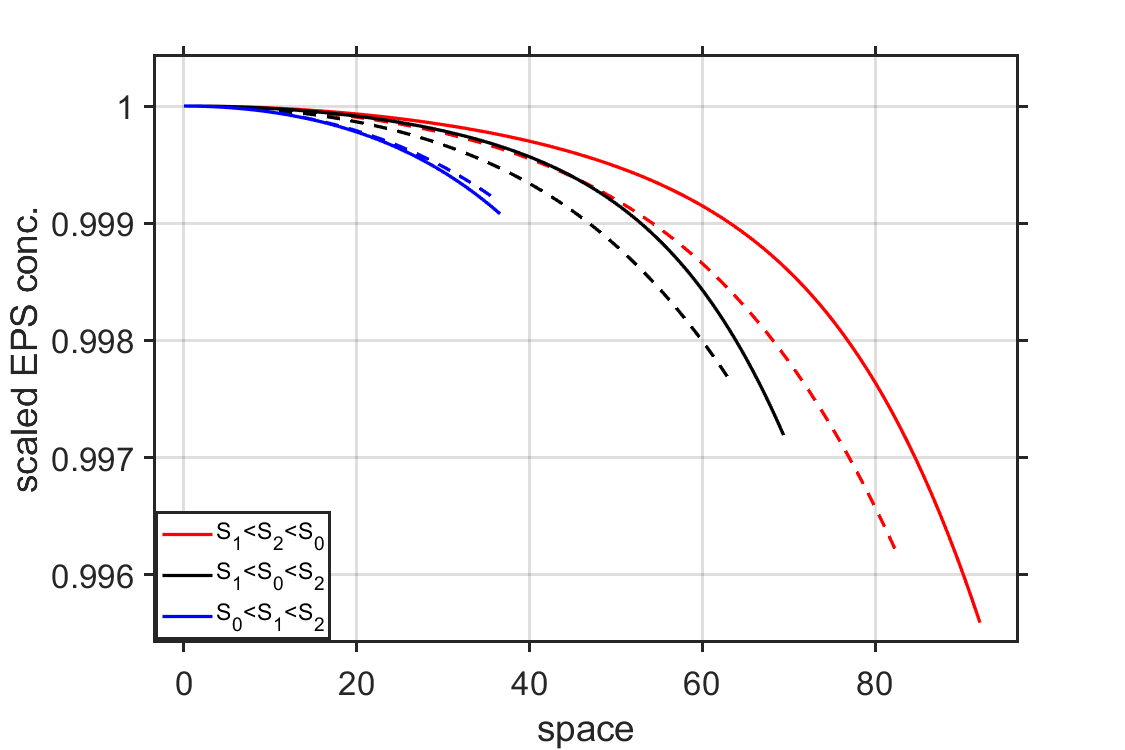}
  \end{subfigure}
 \caption{Effect of varying external nutrient availability ($S_0$) on spatial variation of (a) scaled proliferative bacteria concentration at final time, (b) scaled persister bacteria concentration at final time, (c) scaled dead bacteria concentration at final time, (d) scaled EPS concentration at final time. In each plot, the concentrations are scaled by dividing the value at each spatial point by the concentration at $x=0$. The solid lines refer to the nutrient-dependent phenotype switch model, whereas the dashed lines refer to nutrient-independent phenotype switch model.}
     \label{fig:bbpbdevsxs0scaled}
\end{figure}

Fig~\ref{fig:bbpbdevsxs0scaled} illustrates the spatial distribution of key biomass components—namely, proliferative bacteria, persister bacteria, dead cells, and EPS—across the biofilm at the final simulation time under varying $S_0$. To enhance the visibility of spatial patterns within the biofilm, each concentration profile is normalised by its value at the implant-biofilm boundary ($x=0$). This normalisation facilitates a clearer comparison of relative concentrations and helps identify structural trends across the biofilm depth, extending from the implant interface on the left to the biofilm-bulk fluid boundary on the right. The analysis of these scaled profiles reveals distinct spatial behaviors among the different biomass constituents, offering insights into how each component is distributed in relation to nutrient availability and spatial positioning.

In the cases of proliferative and dead bacteria (Figs.~\ref{fig:bbpbdevsxs0scaled}(a) and (c)), the concentration generally increases as one moves from the implant surface toward the bulk fluid interface. This trend is consistent across both the nutrient-independent phenotype switch and nutrient-dependent phenotype switch models, and under all nutrient conditions examined. The upward trend corresponds to the nutrient gradient, as regions closer to the bulk fluid are richer in nutrients, supporting greater bacterial proliferation and, consequently, more cell death. While the overall shape of the spatial profile remains similar across nutrient levels, the extent of increase becomes more pronounced with higher initial nutrient concentrations, indicating enhanced metabolic activity and turnover in nutrient-rich zones.

For EPS, the spatial distribution displays a decreasing pattern (Fig.~\ref{fig:bbpbdevsxs0scaled}(d)). The highest concentration of EPS is observed near the implant-biofilm interface, where cells are exposed to nutrient limitation. In these nutrient-stressed regions, the EPS production rate exceeds the suppressed bacterial growth rate, leading to higher EPS accumulation. As one moves toward the biofilm-bulk fluid boundary, where nutrient conditions improve, EPS concentrations steadily decline. This inverse relationship between nutrient availability and EPS production is consistently observed across all nutrient levels and in both models, highlighting a stress-induced mechanism for EPS synthesis in the biofilm’s deeper layers.

While the spatial distributions of proliferative bacteria, dead cells, and EPS are qualitatively similar between the two models, the behavior of persister bacteria differs significantly. In the nutrient-independent phenotype switch model,  persister cell concentrations peak near the biofilm-bulk fluid interface—coinciding with regions of high proliferative cell density, since persister formation is solely driven by the abundance of proliferative cells. However, in the nutrient-dependent phenotype switch model, the highest persister concentrations are found near the implant-biofilm boundary. This shift reflects the influence of nutrient limitation on the transition dynamics, with persister formation now being more strongly associated with local environmental stress rather than simply proliferative activity. This contrast underscores the role of nutrient-regulated switching in shaping the spatial organization of bacterial phenotypes within the biofilm.
\end{document}